\pdfoutput=1

\documentclass[11pt,twoside,a4paper,cmspaper,final,collab]{cms-tdr}
\def\svnVersion{166195}\def\cmsCernNo{2011-173}\def\cmsCernDate{\today}\def\cmsMessage{Submitted to Physical Review Letters}

\begin{document}\cmsNoteHeader{EXO-11-038}

\hyphenation{had-ron-i-za-tion}
\hyphenation{cal-or-i-me-ter}
\hyphenation{de-vices}

\RCS$Revision: 97802 $
\RCS$HeadURL: svn+ssh://svn.cern.ch/reps/tdr2/papers/EXO-11-038/trunk/EXO-11-038.tex $
\RCS$Id: EXO-11-038.tex 97802 2012-01-24 18:31:35Z alverson $
\providecommand{\ned}{\ensuremath{n_\textsc{ed}\xspace}}
\providecommand{\MS}{\ensuremath{M_\text{S}\xspace}}
\providecommand{\Mplr}{\ensuremath{{\overline{M}_\mathrm{Pl}}}\xspace}
\newlength\cmsFigWidth
\ifthenelse{\boolean{cms@external}}{\setlength\cmsFigWidth{0.9\columnwidth}}{\setlength\cmsFigWidth{0.6\textwidth}}
\ifthenelse{\boolean{cms@external}}{\providecommand{\CMSleft}{top}}{\providecommand{\CMSleft}{left}}
\ifthenelse{\boolean{cms@external}}{\providecommand{\CMSright}{bottom}}{\providecommand{\CMSright}{right}}
\cmsNoteHeader{EXO-11-038} 

\title{Search for signatures of extra dimensions in the diphoton mass spectrum at the Large Hadron Collider}

\date{\today}

\abstract{
A search for signatures of extra spatial dimensions in the diphoton invariant-mass spectrum has been performed with the CMS detector
at the LHC.  No excess of events above the standard model expectation
is observed using a data sample collected in proton-proton collisions at $\sqrt{s} = 7\TeV$ corresponding to an integrated
luminosity of 2.2\fbinv.  In the context of the large-extra-dimensions model, lower limits are set on the effective
Planck scale in the range of 2.3--3.8 TeV at the 95\% confidence level.  These limits are the most restrictive bounds on
virtual-graviton exchange to date.  The most restrictive lower limits to date are also set on the mass of the first
graviton excitation in the Randall--Sundrum model in the range of 0.86--1.84\TeV, for values of the associated coupling
parameter between 0.01 and 0.10.}

\hypersetup{%
pdfauthor={CMS Collaboration},
pdftitle={Search for signatures of extra dimensions in the diphoton mass spectrum at the Large Hadron Collider},%
pdfsubject={CMS},%
pdfkeywords={CMS, physics, exotica, extra dimensions}}

\maketitle 

Over a decade ago, Arkani-Hamed, Dimopoulos, and Dvali (ADD)~\cite{add,add1} proposed that extra spatial
dimensions could potentially solve the standard model (SM) hierarchy problem~\cite{witten}, which consists of
the observation of the unnatural difference of scales between the gravitational and electroweak theories. They
proposed a scenario whereby the SM is constrained to the common
3+1 space-time dimensions (brane), while gravity is free to propagate throughout a larger
multidimensional space (bulk).  The gravitational flux on the brane is therefore diluted
by virtue of Gauss's Law in the bulk, which relates the fundamental Planck scale $\MD$
to the apparent reduced scale $\Mplr\approx 2\times10^{18}\GeV$ according to the formula $\MD^{\ned+2}=\frac{\Mplr^2}{r^{\ned}}$,
where $r$ and $\ned$ are the size and number of the extra dimensions (ED), respectively.  Postulating a
fundamental Planck scale to be on the order of the electroweak symmetry breaking scale (${\sim}1\TeV$)
results in ED with $r<1\unit{mm}$ for $\ned\ge2$.

Another model of ED that solves the hierarchy problem was proposed by Randall and Sundrum (RS)~\cite{Randall:1999ee}.
In this model, the observed hierarchy is related instead to the warped anti-de Sitter (AdS)
geometry of the ED.  We specifically consider the RS1 model whereby only one
finite ED exists separating two branes, one at each end of the ED.  The geometry of the bulk is based on a slice of
a five-dimensional AdS space with a length $\pi r_\text{c}$, where $r_\text{c}$ is the compactification
radius.  The full metric is
given by $ds^2=e^{-kr_\text{c}y}\eta_{\mu\nu}\,dx^{\mu}dx^{\nu}-r^2_\text{c}dy^{2}$, where Greek indices run over
4-dimensional space-time, $\eta_{\mu\nu}$ is the Minkowski metric tensor, and $0\le y\le \pi$ is the coordinate
along the single ED of radius $r_\text{c}$.  The value of $k$ specifies the curvature scale (or ``warp factor'')
and relates the fundamental Planck scale on one brane to the apparent scale on the other by $\Lambda_\pi=\Mplr e^{-kr_{c}\pi}$.
As a consequence, a value of $k r_\text{c}\sim10$ would provide a natural solution to the hierarchy problem, yielding
$\Lambda_\pi\sim 1\TeV$.

Phenomenologically, the excited gravitons in both models preferentially decay into two gauge bosons, such as photons,
rather than into two leptons, because the graviton has spin 2, and so fermions cannot be produced in an s wave.
In the RS scenario, gravitons appear as well-separated Kaluza--Klein (KK) excitations with masses
and widths determined by the parameters of the RS1 model.  One convenient choice of parameterization
is the mass $M_1$ of the first excitation of the graviton and the dimensionless warp factor
$\tilde{k}\equiv k/\Mplr$, which defines the strength of associated coupling
to the SM fields.  Precision electroweak data constrain $\tilde{k}\gtrsim0.01$, while perturbativity
requirements limit $\tilde{k}\lesssim0.10$~\cite{Davoudiasl:2000wi}.

In the ADD model, the wavefunction of the KK gravitons must satisfy periodic boundary conditions, resulting in
discrete energy levels with modal spacing of the order of the inverse ED size, from 1\unit{meV} to 100\MeV, much
smaller than the spacing in the RS1 model, which is expected to be of the order of 1\TeV.  This effect produces
an apparent continuum spectrum of diphotons, rather than distinct resonances, at high (${\sim}1\TeV$) diphoton
invariant mass $M_{\gamma\gamma}$.

Summing over all KK modes in the ADD scenario results in a divergence in the cross section, so an ultraviolet
(UV) cutoff scale $\MS$ is imposed.  This effective Planck scale is related to---but potentially different
from---the fundamental Planck scale $\MD$.  The precise relationship depends on the UV completion of
the effective theory.  The effects of virtual-graviton production on the differential diphoton cross section are parameterized by
the single variable $\eta_G=\mathcal{F}/\MS^4$, where $\mathcal{F}$ is an order-unity dimensionless
parameter, for which several conventions exist~\cite{CMSdiphoton}.  In this Letter, we set lower limits on $\MS$
in three different conventions: GRW~\cite{GRW}, Hewett~\cite{Hewett}, and HLZ~\cite{HLZ}.

Searches for ED via virtual-graviton effects in the ADD model have been conducted at HERA, LEP, the Tevatron,
and the LHC~\cite{PDG,KCGL}. The most stringent previously published
limits on $\MS$ for $\ned\ge3$ come from the previous measurement in the diphoton channel at the Compact
Muon Solenoid (CMS) experiment~\cite{CMSdiphoton}.  For $\ned=2$, measurements by the D0 experiment
in the dijet~\cite{D0dijet} and diphoton+dielectron~\cite{D0diphoton} channels are more restrictive.  The
most sensitive previous search for RS gravitons was conducted by the ATLAS experiment~\cite{ATLASRSlimit}.  They used
a search in the dilepton final state to exclude $M_1<1.63\TeV$
for $\tilde{k}=0.10$.

In this Letter, we present a search for both non-resonant and resonant diphoton production, in the
ADD and RS1 models, respectively.  We use data corresponding to an integrated luminosity
of 2.2 \fbinv, collected in pp collisions at $\sqrt{s} = 7\TeV$  at the LHC with the CMS
detector between March and August 2011.

The CMS detector~\cite{CMS} is designed to study collisions at the LHC. An all-silicon tracker, an electromagnetic
calorimeter (ECAL), and a hadronic sampling calorimeter (HCAL) are all contained within
a large-bore 3.8\unit{T} superconducting solenoid.  In the central region, the tracker consists
of three radial layers of silicon pixel detectors followed radially by silicon strip detectors.
The finely segmented ECAL has a design resolution for unconverted photons better than $0.5\%$ at energies
exceeding 100\GeV in the barrel ($|\eta|<1.44$).  Here, the pseudorapidity $\eta$ is defined as $-\ln(\tan(\theta/2))$,
where $\theta$ is the polar angle with respect to the direction of the counterclockwise beam.
Beyond the solenoid lie four layers of muon detectors.  The instantaneous
luminosity is measured with a relative uncertainty of 4.5\% using information from forward hadronic
calorimeters~\cite{lumi}.  The two-tiered CMS trigger consists of the level-one trigger, composed of
custom hardware, and the software-based high-level trigger.

Events for the analysis were collected through a diphoton trigger, where each
photon was required to have a transverse energy $\ET\equiv E\sin\theta$ of at least
33, 55, or 60\GeV, depending on the instantaneous luminosity.  We require that an event be
consistent with a pp collision and have at least one
well-re\-con\-struct\-ed primary vertex~\cite{CMSVertex}.  We then reconstruct photons with
$\ET>70$\GeV in the ECAL barrel by clustering electromagnetic energy depositions.
Electrons that do not originate from photon conversions are suppressed by using information from the pixel
detector to associate tracks and ECAL clusters compatible with an electron hypothesis.  The probability
of misidentification of an electron as a photon is approximately 3\% resulting in a negligibly small
contribution to the diphoton spectrum in the signal region.

Hadronic jets can be misidentified as photons when their leading hadron is an energetic $\pi^0$
or $\eta$ meson.  We reduce the misidentification rate from this source by placing the same restrictions on the isolation
as in the previous analysis for this channel~\cite{CMSdiphoton}. These restrictions limit the total
transverse energy because of tracks and calorimeter depositions near the photon cluster. Restrictions
on the shower-shape variable $\sigma_{\eta\eta}$, which is a modified second moment of the
electromagnetic energy cluster about its mean $\eta$ position~\cite{CMSQCDPhoton}, also suppress
hadronic misidentification.  Topological and timing criteria suppress anomalous signals present in the
ECAL~\cite{EGM-10-006}. Diphoton events are selected in which $M_{\gamma\gamma}>140\GeV$.

The photon reconstruction and identification efficiency is determined in Monte Carlo (MC) simulation and
corrected using a data-to-MC scale factor of $1.005\pm0.034$ derived from studying $\cPZ\to\Pep\Pem$ events. The measured efficiency for
a single $\ET>70$\GeV photon with $|\eta|<1.44$ is $(87.4\pm5.4)\%$ and depends only weakly on the
\ET and $\eta$ of the photon, and the number of extra collisions present in the event.  The
systematic uncertainty bounds the variation as a function of these variables, the most significant of which is the
number of extra collisions.  We reweight the simulation to give the same reconstructed primary-vertex distribution
(on average 6--8 vertices) as observed in the data.  We determine the corresponding diphoton
reconstruction and identification efficiency $(76.4\pm9.6)\%$ by squaring the single-photon efficiency.

The simulation of ED in the ADD model is performed using version 1.3.0 of the \SHERPA~\cite{Sherpa} MC generator.
The simulation includes both SM diphoton production and signal diphoton production via virtual-graviton
exchange in order to account for the interference effects between the SM and ADD processes. The leading-order (LO) \SHERPA cross
sections are multiplied by a constant next-to-leading-order (NLO) $K$ factor of $1.6 \pm 0.1$, a value that
represents an updated calculation for $\sqrt(s)=7$~TeV by the authors of Refs.~\cite{ggNLO, EDNLO}.  The systematic
uncertainty on the signal $K$ factor reflects the approximate variation of the $K$ factor over a large region
of the model parameters; it is not intended to account for the theoretical uncertainty.
The cross sections in the simulation are conservatively set to zero for $\sqrt{\hat{s}}>\MS$
because the theory becomes non-perturbative for larger values of $\sqrt{\hat{s}}$.  Introducing this sharp
truncation reduces the upper limits on $\MS$ by a few percent.

The simulation of RS-graviton production is performed using version 6.424 of the \PYTHIA~\cite{pythia} MC program.
The signal cross section is scaled by a mass-dependent NLO $K$ factor~\cite{ggNLO, EDNLO},
which ranges from 1.6 to 1.8 as a function of $M_{\gamma\gamma}$ and for different values of $\tilde{k}$.
The \textsc{cteq6l1}~\cite{CTEQ} parton distribution functions (PDF)
are used in the simulation of both the ADD and RS models, and a $1.5\%$ relative uncertainty on the
signal acceptance is included by measuring its dependence on the choice of PDF and its uncertainties.

Optimization of the event selection is done separately for both ADD and RS scenarios.  The signal
in both cases is predominantly at central values of $\eta$, while the high-$M_{\gamma\gamma}$ SM background dominates the signal in the
forward region; therefore, we restrict ourselves to photons located in the ECAL barrel only.  In the ADD scenario,
we find that the optimal region for the search, based on the expected signal significance, is $M_{\gamma\gamma}>900$~\GeV.
This choice of selection depends weakly on the model parameters.

In the search for RS gravitons, a fixed window is selected about the $M_1$ mass point of interest.  Because the signal
shapes deviate from Gaussian distributions, we define an effective measure of the signal width $\sigma_\text{eff}$ as the
half-width of the narrowest mass interval containing 68\% of the signal from simulation.  The value of $\sigma_\text{eff}$
ranges from 6 to 21~GeV for RS gravitons with $M_1$ between 500 and 2000~GeV and $\tilde{k}=0.01$.  The dependence on $M_1$
is linear and also increases with $\tilde{k}$ ($\sigma_\text{eff}=42$~GeV for $M_1=2$~TeV and $\tilde{k}=0.10$).  A window
is then formed about the resonance mean of size $\pm5\,\sigma_\text{eff}$ in the data.  This window contains 96--97\% of
the signal acceptance for all mass points considered in this analysis, and the detector resolution is negligible with
respect to the window size.  This choice of the window maximizes the signal acceptance and analysis sensitivity in the
case of small backgrounds.

Backgrounds from the misidentification of a hadronic jet as a photon are small in the signal region but contribute to the
low-$M_{\gamma\gamma}$ region.  Two such sources of backgrounds from isolated-photon misidentification
are considered: multijet production and prompt single-photon ($\gamma+\text{jet}$) production.  In particular, we measure on
a background-dominated sample a misidentification rate, defined as the ratio of the number of isolated
photon candidates to non-isolated photon-like objects.  These photon-like objects are reconstructed as
photons but fail one of the isolation or shower-shape criteria; therefore, the samples corresponding to
numerator and denominator are mutually exclusive, and prompt photons have a negligibly small contribution to the
denominator.
The misidentification rate is measured in a photon-triggered sample in bins of photon(-like) candidate $E_{\rm T}$, but the
objects used in the measurement are required to be well-separated from the triggered object to avoid a
trigger-induced bias.

Because the background-dominated sample in which we measure the misidentification rate may contain some genuine,
isolated photons that ``contaminate'' the numerator of the misidentification rate, we correct for this on a
bin-by-bin basis.  The $\sigma_{\eta\eta}$ requirement is released and the numerator sample is fit
for the fraction of prompt photons using one-dimensional probability density histograms (``templates'') in
$\sigma_{\eta\eta}$.  The signal template is constructed from MC simulation, and the background template is constructed
from reconstructed photons that fail one or more of the isolation criteria.  The measured misidentification rate falls
from 7\% at $\ET=70$ GeV to 2\% at $\ET=120$\GeV.  We apply a 20\% systematic uncertainty to the rate derived
from the variation of the misidentification rate measured in a jet-triggered sample.

The multijet and $\gamma+\text{jet}$ backgrounds to the reconstructed diphoton spectrum are estimated by using the
misidentification rate to extrapolate from two background-dominated reference regions, both selected with the same
diphoton trigger as the primary signal sample.  One region includes events with only one isolated photon, but
one or more non-isolated photons.  The other region includes events with no isolated photons, but two or more
non-isolated photons.  The diphoton trigger is sufficiently inclusive that the regions are unaffected by the trigger
selection.  By applying the prompt-photon misidentification rate to these two reference regions, we predict the
$\gamma+\text{jet}$ and multijet backgrounds in the signal region.

The SM diphoton background dominates the signal region.  The expected number of background events due to this process is computed by
rescaling the prediction from \PYTHIA with an NLO $K$ factor that varies with $M_{\gamma\gamma}$.  The NLO prediction
is calculated with the \textsc{diphox}+ \textsc{gamma2mc}~\cite{diphoxPaper,gamma2MCPaper} generators, which take into
account the fragmentation processes in which the photons can come from the collinear fragmentations of hard partons.  A separate
analysis by CMS has also demonstrated good agreement with the NLO prediction at low $M_{\gamma\gamma}\lesssim300\GeV$~\cite{CMS:2011qt}.
The sub-leading-order gluon-fusion box diagram is included as a part of the \textsc{pythia} calculation because of its large contribution
at the LHC energy, although its effects are small at high $M_{\gamma\gamma}$.  The $K$ factor varies between 1.7 and 1.1 from low
to high $M_{\gamma\gamma}$.  A systematic uncertainty of 15\% on the value of the $K$ factor is determined by examining
the PDF uncertainties and variation of the renormalization and factorization scales.

\begin{figure}[!htbp]
\centering
\includegraphics[width=\cmsFigWidth]{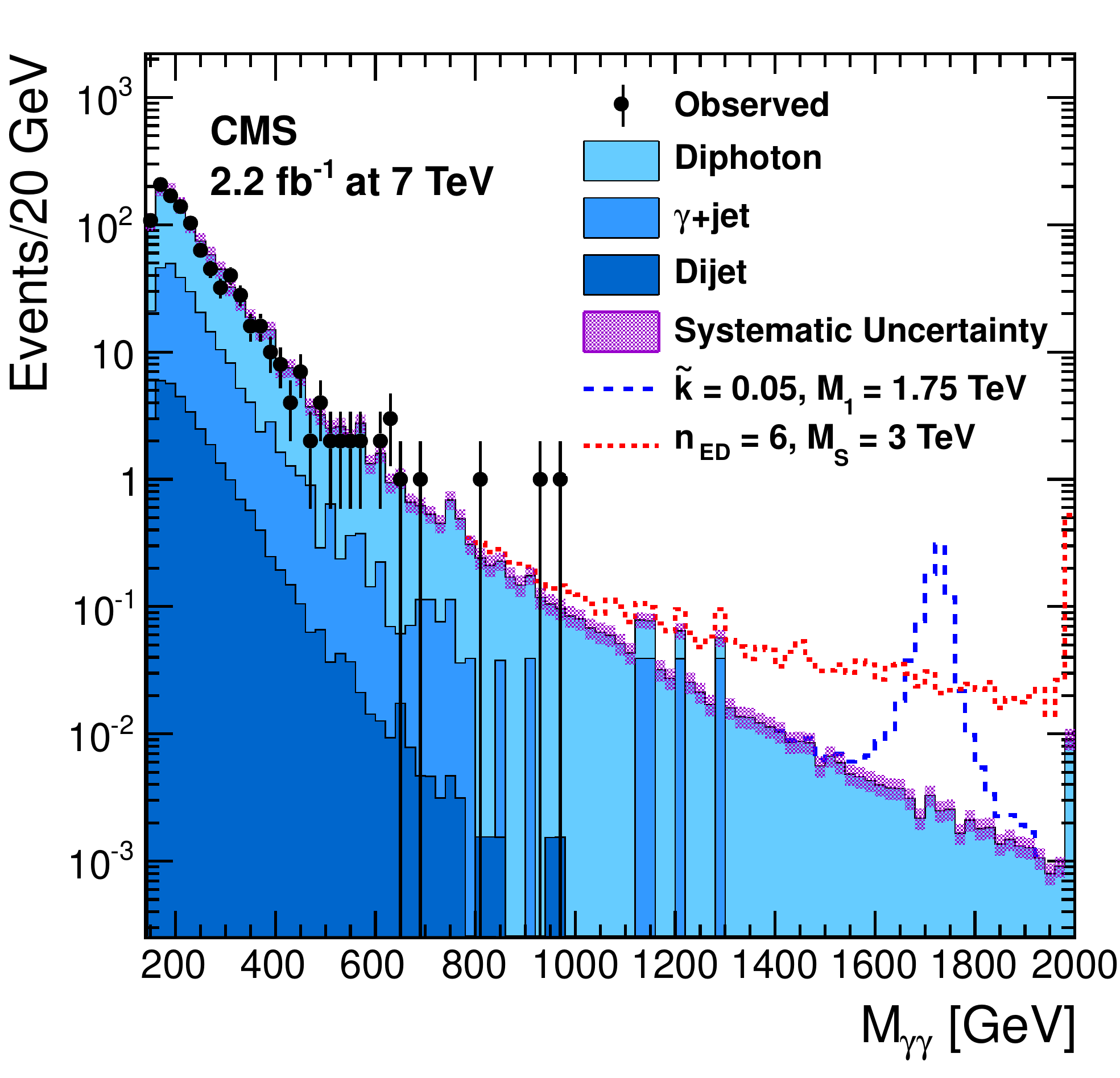}
\caption{Observed event yields (points with error bars) and background expectations (filled solid histograms) as a function of the
diphoton invariant mass. Photons are required to be isolated, with $\ET>70$\GeV and $|\eta|<1.44$. The shaded band around
the background estimation corresponds to the average systematic uncertainty over the spectrum.  The precise per-bin uncertainty is not provided for the sake of clarity.  The last bin includes the sum of all
contributions for $M_{\gamma\gamma}>2.0\TeV$.  The simulated distributions for two, non-excluded signal hypotheses are shown for
comparison as dotted (ADD) and dashed (RS) lines.\label{fig:results_invmass}}
\end{figure}

Figure~\ref{fig:results_invmass} shows the invariant mass distribution of the selected events, together
with the estimated distributions for each of the backgrounds.  Table~\ref{table:eventcontent} presents
the observed number of events in the data and the predicted number of background events in different ranges in $M_{\gamma\gamma}$
and corresponds directly to Fig.~\ref{fig:results_invmass}.  The last column corresponds to the signal
region for the ADD search.  We find that the observed data are consistent with the background estimate throughout the
$M_{\gamma\gamma}$ spectrum and do not show an excess of events, neither resonant nor non-resonant.

\begin{table}[!htbp]
 \centering
\caption{Observed event yields and background expectations for different reconstructed diphoton invariant-mass ranges.
Full systematic uncertainties are included.\label{table:eventcontent}}
 \begin{tabular}{ccccc}
 \hline
 \hline
 \multirow{2}{*}{Process} & \multicolumn{4}{c}{Diphoton Invariant Mass Range [TeV]} \\
                          & [0.14, 0.2] & [0.2, 0.5] & [0.5, 0.9] & $>0.9$ \\
\hline
 Multijet             &  15 $~\pm $   6  &   17 $~\pm $  7   & 0.2 $~\pm $ 0.1  & 0.003 $~\pm $ 0.001 \\
 $\gamma+\text{jet}$  &  102 $~\pm $ 15  &   124 $~\pm $ 18  & 2.5 $~\pm $ 0.4  & 0.19 $~\pm $ 0.04   \\
 Diphoton             &  372 $~\pm $ 70  &   414 $~\pm $ 78  & 16.9 $~\pm $ 3.2  & 1.3 $~\pm $ 0.3 \\
\hline
 Backgrounds    &  489 $~\pm $ 73  &   555 $~\pm $ 81  & 19.6 $~\pm $ 3.2  & 1.5 $~\pm $ 0.3 \\
 Observed      &  484         &   517            & 16               & 2              \\
 \hline
 \hline
\end{tabular}
\end{table}

To set limits on virtual-graviton exchange in the ADD scenario, we compare the number of observed and
expected events in the signal region ($M_{\gamma\gamma}>0.9\TeV$) and set 95\% confidence level (CL) upper limits on the quantity
$S\equiv(\sigma_\text{total}-\sigma_\text{SM})\times\mathcal{B}\times\mathcal{A}$,
where $\sigma_\text{total}$ represents the total diphoton production cross section (including signal,
SM, and interference effects), and $\sigma_\text{SM}$ represents the SM diphoton production cross section.
The signal branching fraction to diphotons is indicated by $\mathcal{B}$ and the signal acceptance by $\mathcal{A}$.
We use the CL$_\text{s}$ technique~\cite{CLS1, CLS2} to compute the limits with a likelihood constructed from the Poisson
probability to observe $N$ events, given $S$, the signal efficiency $(76.4\pm9.6)\%$, the expected number
of background events ($1.5\pm0.3$), and the integrated luminosity $\mathcal{L}=(2.2\pm0.1)\fbinv$~\cite{lumi}.
The variation of the $K$ factor is included in the statistical analysis as an uncertainty on the signal yield.

\begin{figure}[!htbp]
\centering
\includegraphics[width=0.4\textwidth]{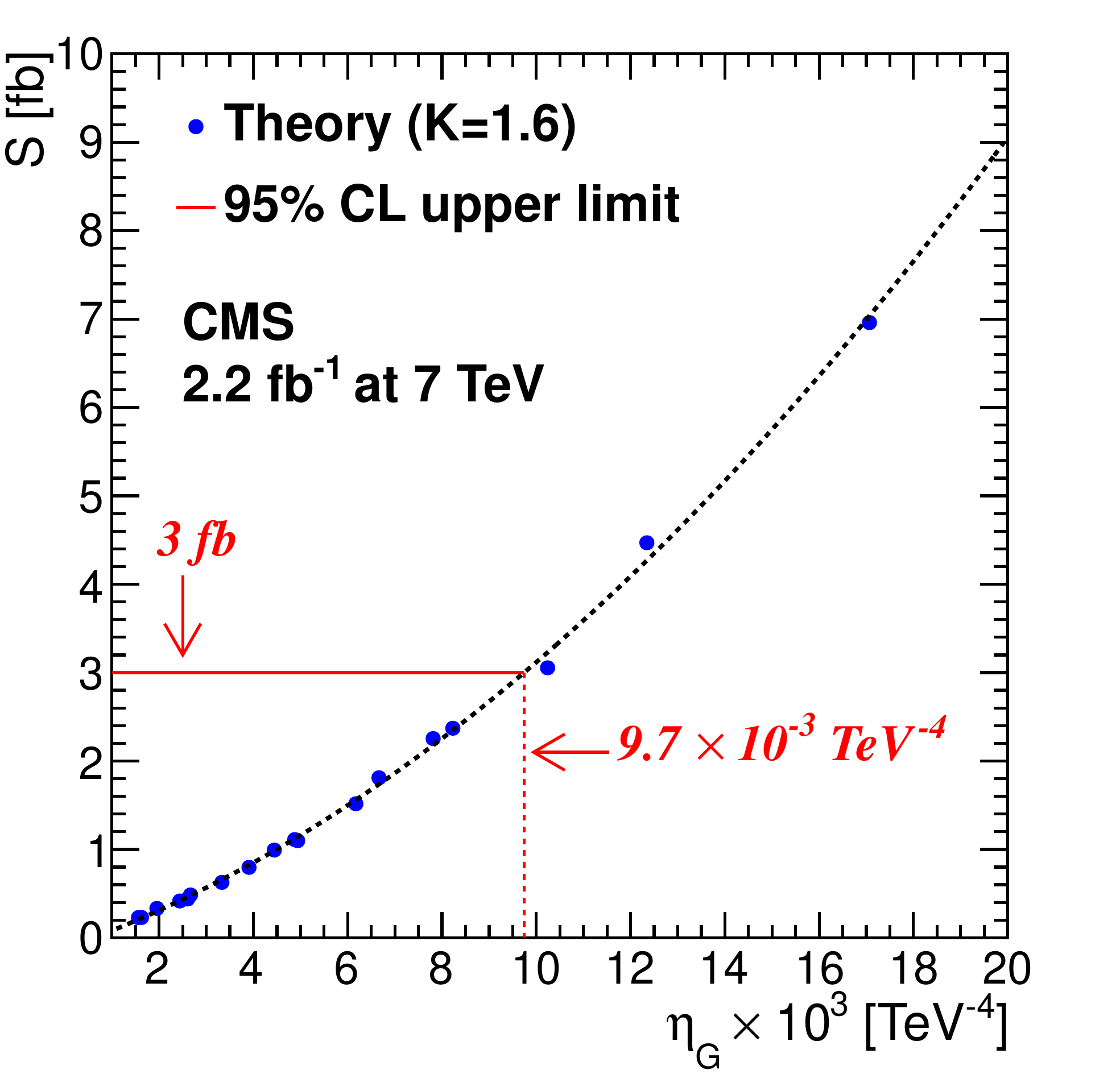}
\includegraphics[width=0.4\textwidth]{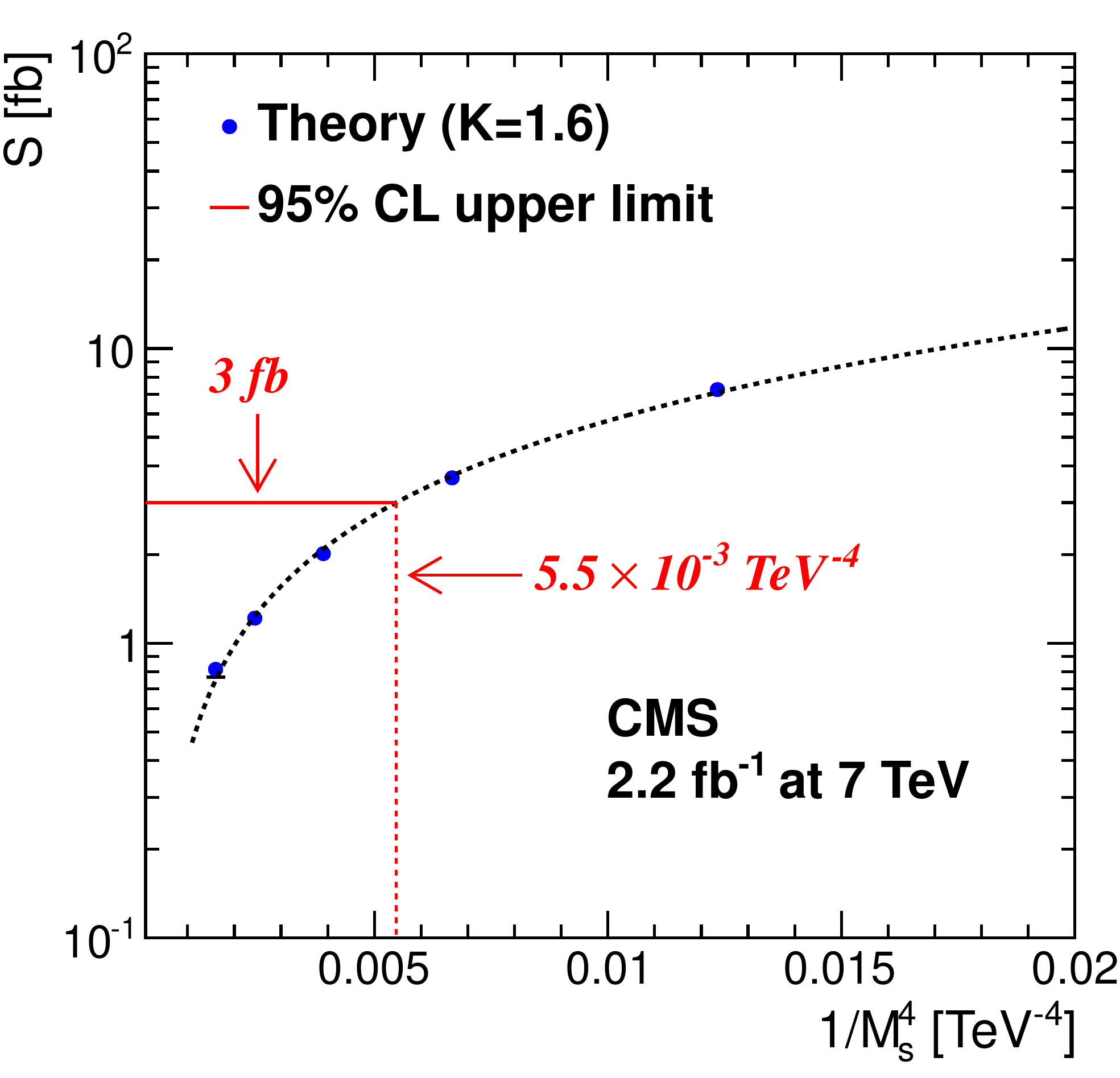}
\caption{Signal cross section $S$ parameterization as a function of the strength of the ED effects, $\eta_G$ (\CMSleft) and as a function of $1/\MS^4$ for the HLZ $\ned=2$ case (\CMSright).\label{fig:cs-limit}}
\end{figure}

The observed (median expected) 95\% CL upper limit on $S$ is 3.0\unit{fb} (2.7\unit{fb}).  For the HLZ $\ned=2$ case, we parameterize $S$ directly as a
smooth function of $1/\MS^4$.  For all other conventions, $S$ is parameterized as a function of the parameter $\eta_G$,
as in Ref.~\cite{CMSdiphoton}. The observed 95\% CL limit, together with the signal parameterization, are shown in Fig.~\ref{fig:cs-limit}. The intersection of the cross-section limit with the parameterized curve determines the 95\% CL upper limit on the parameter
$\eta_G$. As seen from the plot, these upper limits on $S$ correspond to upper limits of $\eta_G\le 0.0097\TeV^{-4}$
and $1/\MS^4\le 0.0055\TeV^{-4}$.  The upper limits on $\eta_G$ are equated to lower limits on $\MS$
and are shown in Table~\ref{table:limits}.

\begin{table*}[!htbp]
\centering
\caption{The $95\%$ CL lower limits on $\MS$ (in \TeV) in the GRW, Hewett, and HLZ conventions for two values of the ADD
signal $K$ factor, 1.0 and $1.6\pm0.1$.  All limits are computed with a signal cross section truncated to zero for
$\sqrt{\hat s}>\MS$,  where $\sqrt{\hat s}$ is the center-of-mass of the partonic collision.  The limits are presented
for both positive and negative interference in the Hewett convention and for $\ned=2$--7 in the HLZ convention.  The median
expected lower limits are given in parenthesis.
\label{table:limits}}
\begin{tabular}{cccccccccc}
\hline\hline
\multirow{2}{*}{$K$} & \multirow{2}{*}{GRW} & \multicolumn{2}{c}{Hewett} & \multicolumn{6}{c}{HLZ} \\
& & pos. & neg. & $\ned=2$ & $\ned=3$ & $\ned=4$ & $\ned=5$ & $\ned=6$ & $\ned=7$ \\
\hline
\multirow{2}{*}{1.0} & 2.94 & 2.63 & 2.28 & 3.29 & 3.50 & 2.94 & 2.66 & 2.47 & 2.34 \\
                     & (2.99) & (2.67) & (2.31) & (3.37) & (3.56) & (2.99) & (2.71) & (2.52) & (2.38) \\
\hline
\multirow{2}{*}{$1.6\pm0.1$} & 3.18 & 2.84 & 2.41 & 3.68 & 3.79 & 3.18 & 2.88 & 2.68 & 2.53 \\
                             & (3.24) & (2.90) & (2.44) & (3.77) & (3.85) & (3.24) & (2.93) & (2.73) & (2.58) \\
\hline\hline
\end{tabular}
\end{table*}

\begin{figure}[!htbp]
\centering
\includegraphics[width=\cmsFigWidth]{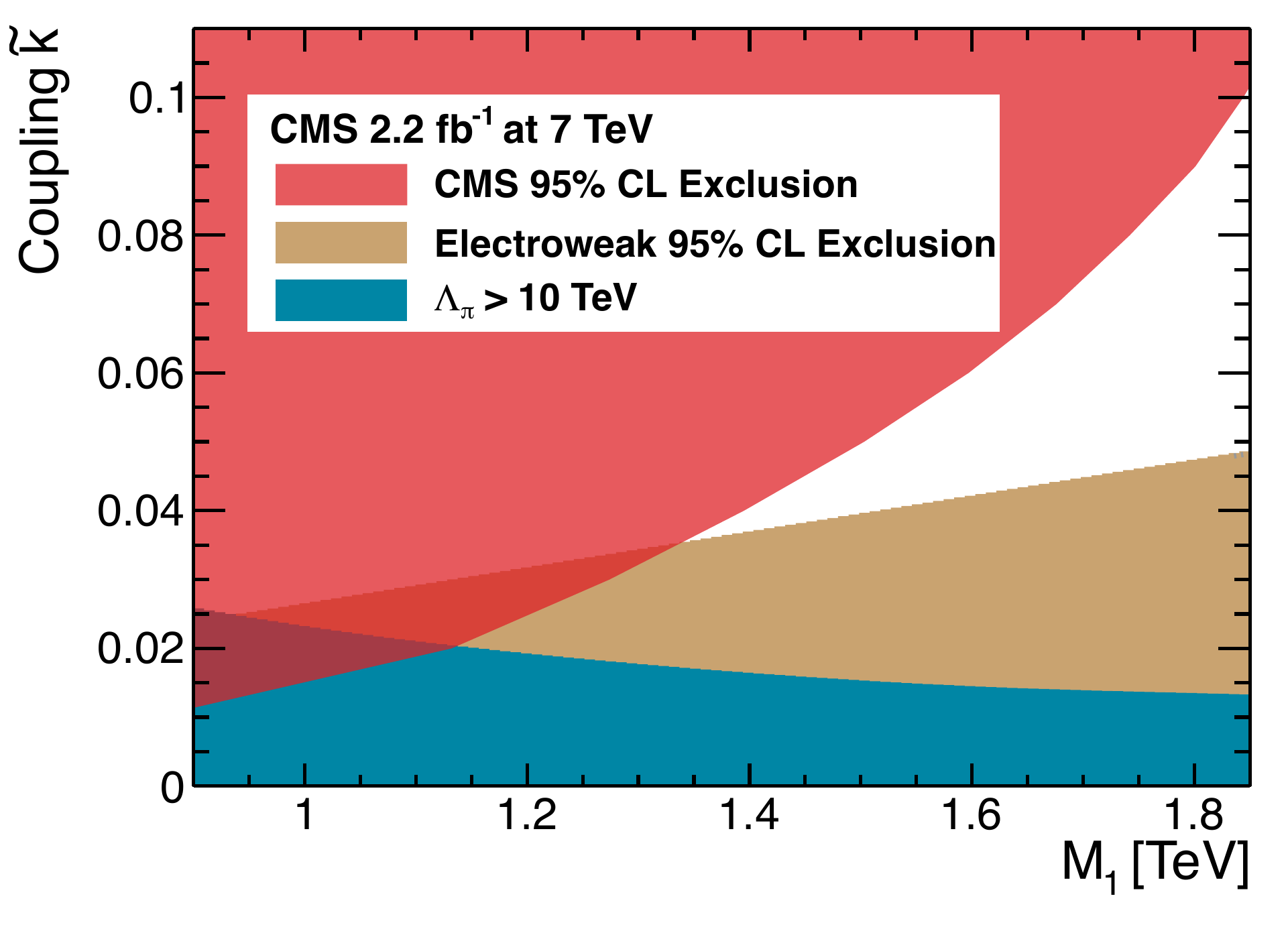}
\caption{The 95\% CL exclusion region for the RS1 graviton model in the $M_1$--$\tilde{k}$ plane.  The expected
limits coincide very closely with the measured limits and so are not shown in the figure.  Also shown
are bounds due to electroweak constraints and naturalness ($\Lambda_{\pi}>10\TeV$).
Perturbativity requirements bound $\tilde{k}\lesssim0.10$.}
\label{fig:RSLimit2}
\end{figure}

For the RS scenario, the same limit-setting calculation is performed, but in a bounded window in $M_{\gamma\gamma}$.
Figure~\ref{fig:RSLimit2} shows the excluded regions in the $M_1$--$\tilde{k}$ plane.
Also shown are bounds due to precision electroweak measurements and to naturalness arguments~\cite{Davoudiasl:2000wi}.
Table~\ref{table:RSLimits} presents the 95\% CL lower limits on the graviton mass $M_1$ for different
values of $\tilde{k}$.  The median expected lower limits coincide within a few GeV of the observed limits.

\begin{table*}[!htbp]
\begin{center}
\caption{The 95\% CL lower limits on $M_1$ for given values of the coupling parameter, $\tilde{k}$.  For $\tilde{k}<0.03$,
masses above the presented limits are excluded by electroweak and naturalness constraints.  The median expected lower limits
are numerically the same for the presented precision except for the $\tilde{k}=0.01$ case, for which the expected lower
limit on $M_1$ is 0.84~TeV.}
\label{table:RSLimits}
\begin{tabular}{ccccccccccc}
\hline\hline
$\tilde{k}$ \rule{0pt}{2.6ex} & 0.01 & 0.02 & 0.03 & 0.04 & 0.05 & 0.06 & 0.07 & 0.08 & 0.09 & 0.10 \\
$M_1$ [TeV] & 0.86 & 1.13 & 1.27 & 1.39 & 1.50 & 1.59 & 1.67 & 1.74 & 1.80 & 1.84 \\
\hline\hline
\end{tabular}
\end{center}
\end{table*}

In summary, we have performed a search for extra spatial dimensions leading to enhanced
resonant or non-resonant diphoton production in proton-proton collisions at a center-of-mass energy
of 7\TeV at the LHC. Using a data sample corresponding to an integrated luminosity of 2.2\fbinv
recorded by the CMS experiment, we observe no excess in diphoton production above the rate predicted
from SM background sources. Values of the effective Planck scale $\MS$ less than 2.3--3.8\TeV
are excluded at 95\% CL for ADD models.  We also exclude at 95\% CL resonant graviton production in
the RS1 model with values of $M_1$ less than 0.86--1.84\TeV depending on the normalized coupling
strength, $\tilde{k}$.   We present limits on both the ADD and RS1 models of extra dimensions in the
diphoton final state that extend those observed at the D0 experiment~\cite{D0diphoton}, as well as those set
previously by the CMS~\cite{CMSdiphoton} and ATLAS~\cite{ATLASRSlimit} experiments.

We thank M.C.~Kumar, P.~Mathews, V.~Ravindran, and A.~Tripathi for the calculation of NLO $K$ factors used in this Letter.
We wish to congratulate our colleagues in the CERN accelerator departments for the excellent performance of the LHC machine.
We thank the technical and administrative staff at CERN and other CMS institutes, and acknowledge support from: FMSR (Austria); FNRS and FWO (Belgium); CNPq, CAPES, FAPERJ, and FAPESP (Brazil); MES (Bulgaria); CERN; CAS, MoST, and NSFC (China); COLCIENCIAS (Colombia); MSES (Croatia); RPF (Cyprus); Academy of Sciences and NICPB (Estonia); Academy of Finland, MEC, and HIP (Finland); CEA and CNRS/IN2P3 (France); BMBF, DFG, and HGF (Germany); GSRT (Greece); OTKA and NKTH (Hungary); DAE and DST (India); IPM (Iran); SFI (Ireland); INFN (Italy); NRF and WCU (Korea); LAS (Lithuania); CINVESTAV, CONACYT, SEP, and UASLP-FAI (Mexico); MSI (New Zealand); PAEC (Pakistan); SCSR (Poland); FCT (Portugal); JINR (Armenia, Belarus, Georgia, Ukraine, Uzbekistan); MST, MAE and RFBR (Russia); MSTD (Serbia); MICINN and CPAN (Spain); Swiss Funding Agencies (Switzerland); NSC (Taipei); TUBITAK and TAEK (Turkey); STFC (United Kingdom); DOE and NSF (USA).

\bibliography{auto_generated}   
\cleardoublepage \appendix\section{The CMS Collaboration \label{app:collab}}\begin{sloppypar}\hyphenpenalty=5000\widowpenalty=500\clubpenalty=5000\textbf{Yerevan Physics Institute,  Yerevan,  Armenia}\\*[0pt]
S.~Chatrchyan, V.~Khachatryan, A.M.~Sirunyan, A.~Tumasyan
\vskip\cmsinstskip
\textbf{Institut f\"{u}r Hochenergiephysik der OeAW,  Wien,  Austria}\\*[0pt]
W.~Adam, T.~Bergauer, M.~Dragicevic, J.~Er\"{o}, C.~Fabjan, M.~Friedl, R.~Fr\"{u}hwirth, V.M.~Ghete, J.~Hammer\cmsAuthorMark{1}, M.~Hoch, N.~H\"{o}rmann, J.~Hrubec, M.~Jeitler, W.~Kiesenhofer, A.~Knapitsch, M.~Krammer, D.~Liko, I.~Mikulec, M.~Pernicka$^{\textrm{\dag}}$, B.~Rahbaran, H.~Rohringer, R.~Sch\"{o}fbeck, J.~Strauss, A.~Taurok, F.~Teischinger, C.~Trauner, P.~Wagner, W.~Waltenberger, G.~Walzel, E.~Widl, C.-E.~Wulz
% ...
\vskip\cmsinstskip
\dag:~Deceased\\
1:~~Also at CERN, European Organization for Nuclear Research, Geneva, Switzerland\\
% ...

\end{sloppypar}
\end{document}